\documentstyle[12pt]{article}
\normalsize
\def\sp{~~~~~}

\let\oldtheequation=\theequation
\def\doteqs#1{\setcounter{equation}{0}
            \def\theequation{{#1}.\oldtheequation}}

\newcounter{sxn}
\def\sx#1{\addtocounter{sxn}{1} \bigskip\medskip \goodbreak \noindent{\large\bf
\centerline{\thesxn.~~#1}} \nobreak \medskip}
\def\sxn#1{\sx{#1} \doteqs{\thesxn}}

\newcounter{axn}

\def\br{\begin{thebibliography}{abc}}
\def\er{\end{thebibliography}}

\date{}

\begin{document}
\bibliographystyle{unsrt}
\footskip 1.0cm
\thispagestyle{empty}
\begin{flushright}
SU-4228-487\\
INFN-NA-IV-91/12\\
UAHEP 917\\
October 1991\\
\end{flushright}
%\vspace*{5mm}
\begin{center}{\LARGE CONFORMAL EDGE CURRENTS\\
         IN\\
       CHERN-SIMONS THEORIES\\}
\vspace*{10mm}
{\large A. P. Balachandran,$^{1}$
          G. Bimonte $^{1,2}$ \\
          K. S. Gupta,$^{1}$
          A. Stern $^{2,3}$ \\ }
\newcommand{\bc}{\begin{center}}
\newcommand{\ec}{\end{center}}
\vspace*{10mm}
 1){\it Department of Physics, Syracuse University,\\
Syracuse, NY 13244-1130, USA}.\\
\vspace*{5mm}
 2){\it Dipartimento di Scienze Fisiche dell' Universit\`a di Napoli,\\
    Mostra d'Oltremare pad. 19, 80125 Napoli, Italy}.\\
\vspace*{5mm}
 3){\it Department of Physics, University of Alabama, \\
Tuscaloosa, Al 35487, USA.}\ec

\vspace*{10mm}
\normalsize
\centerline{\bf ABSTRACT}
\vspace*{7mm}

We develop elementary canonical methods for the quantization of abelian and
nonabelian Chern-Simons actions using well known ideas in gauge theories and
quantum gravity.  Our approach does not involve choice of gauge or clever
manipulations of functional integrals.  When the spatial slice is a disc, it
yields Witten's edge states carrying a representation of the  Kac-Moody
algebra.  The canonical expression for the generators of diffeomorphisms on the
boundary of the disc are also found, and it is established that they are the
Chern-Simons version of the Sugawara construction. This paper is a prelude to
our future publications on edge states, sources,
vertex operators, and their spin and statistics in 3d and 4d topological field
theories.

\newpage

\newcommand{\be}{\begin{equation}}
\newcommand{\ee}{\end{equation}}

\baselineskip=24pt
\setcounter{page}{1}
\sxn{INTRODUCTION}

\indent
The Chern-Simons or CS action describes a three-dimensional field theory of a
connection $A_\mu$. In the absence of sources, the  field equations
require $A_\mu$
to be a zero curvature field and hence to be a pure gauge in simply connected
spacetimes.  As the dynamics is  gauge invariant as well, it would appear that
the CS action is an action for triviality in these spacetimes.

Such a conclusion however is not always warranted.  Thus, for instance, it is
of frequent interest to consider the CS action on a disc $D$ $\times {\bf R}^1$
(${\bf R}^{1}$ accounting for time)
and in this case, as first emphasized by Witten
 \cite{witt}, it is possible to
contemplate a quantization which eliminates degrees of freedom only in the
interior of $D$.  In such a scheme, then, gauge transformations relate
equivalent fields only in the interior of $D$ whereas on the boundary $\partial
D$, they
play a role more akin to global symmetry transformations.  The residual states
localized on the circular boundary $\partial D$ are the CS edge states.
As they are associated with gauge transformations
on $\partial D$ = the circle S$^1$, it is natural to expect  that the
loop or the Kac-Moody group \cite{godd} of the gauge group will play a role in
their description, the latter being a central extension of the former. Witten
\cite{witt} in fact outlined an argument to show that the edge states form a
conformal family carrying a representation of the Kac-Moody group.

Subsequent developments in the quantum theory of CS action have addressed both
its formal \cite{bos,bal,guad,lee,moore}
and its physical \cite{zhang,blok,mor} aspects.
As regards the former, methods have been invented and refined for its fixed
time quantization \cite{bos,bal,guad}
and for the treatment of its functional
integral \cite{guad,lee,moore}.  They yield Witten's
results and extend them as well.  An important achievement of all this research
beginning in fact with Witten's work is the reproduction of a large class of
two-dimensional (2d) conformal field theories (CFT's) from 3d CS theories.

\indent
There have been equally interesting developments which establish the
significance of the CS interaction for 2d condensed matter systems which go
beyond phase transition phenomena described by CFT's \cite{godd}.
  It is now well
appreciated for instance that the edge states of the Fractional Quantum Hall
Effect (FQHE) are well described by the CS theory and its variants
 \cite{zhang,blok} and that
it is of fundamental importance in the theory of fractional statistics
 \cite{mor}.
Elsewhere, we will also describe its basic role in the theory of London
equations of 2d superconductors.

In this paper, we develop a canonical quantization of the CS action assuming
for simplicity that spacetime is a solid cylinder $D \times {\bf R}^{1}$.  A
notable merit of
our approach is that it avoids making a gauge choice or delicate
manipulations of functional integrals. It is furthermore
based on ideas which are standard in
field theories with constraints such as QCD or quantum gravity \cite{marmo} and
admits easy generalizations, for example, to 4d gauge theories.  In subsequent
papers, we will extend this approach to certain gauge field theories (including
the CS theory) with sources.  We will establish that an anyon or a Laughlin
quasiparticle is not just a single particle, but is in reality a conformal
family (a result due to Witten \cite{witt}) and derive similar results in
four dimensions.  Simple considerations concerning spin and statistics of these
sources will also be presented using basic ideas of European
schools \cite{buch}  on
``fields localized in space--like cones'' and generalizing them somewhat. A
brief account of our work has already appeared elsewhere \cite{bim}.

In Section 2, we outline  a canonical formalism for the U(1) CS action on $D
\times {\bf R}^{1}$ and
its relation to certain old ideas in gauge theories or gravity.  The
observables are shown to obey an algebra isomorphic to the U(1) Kac--Moody
algebra on a circle \cite{godd}. The classical canonical expression for the
diffeomorphism (diffeo) generators on $\partial D$ are also found.

In Section 3, the observables are Fourier analyzed on $\partial D$.  It is
then discovered that the CS diffeo generators are weakly the same as those
obtai
ned
by the Sugawara construction \cite{godd}.  Quantization is then carried out in
a
conventional way to find that the edge states and their observables describe a
central charge 1 conformal family \cite{godd}. We next briefly illustrate our
techniques by quantizing a generalized version of the CS action
which has proved important in the theory of FQHE \cite{blok}.

The paper concludes with Section 4 which outlines the nonabelian version of the
foregoing considerations.

\sxn{THE CANONICAL FORMALISM}

\indent
The U(1) CS action on the solid cylinder $D \times {\bf R}^{1}$ is
\begin{equation}
S=\frac{k}{4\pi} \int_{D \times {\bf R}^1} A dA,~~ A = A_\mu dx^\mu,
{}~~AdA \equiv A\wedge dA
\end{equation}
where $A_{\mu}$ is a real field.

The action S is invariant under diffeos of the solid cylinder and does not
permit a natural choice of a time function.  As time is all the same
indispensable in the canonical approach, we arbitrarily choose a time function
denoted henceforth by $x^0$. Any constant $x^0$ slice of the solid
cylinder is then the disc $D$ with
coordinates $x^1$, $x^2$.

It is well known that the phase space of the action $S$ is
described by the equal
time Poisson brackets (PB's)
\be
\left \{A_{i}(x),A_{j}(y)\right \}=\epsilon_{ij}\frac{2\pi}{k}\delta^{2}(x-y)
 ~~ {\rm for}~ i,j=1,2 ,
\;\;\;\;\epsilon_{12}=-\epsilon_{21}=1
\ee
(using the convention $\epsilon^{012} = 1$ for the Levi-Civita symbol)
and the constraint
\be
\partial_{i}A_{j}(x) - \partial_{j}A_{i}(x) \equiv F_{ij}(x) \approx 0
\ee
where $\approx$ denotes weak equality in the sense of Dirac \cite{marmo}.
All fields
are evaluated at the same time $x^0$ in these equations, and this will continue
to be the case when dealing with the canonical formalism or quantum operators
in the remainder of the paper.  The connection $A_0$ does not occur as a
coordinate of this phase space.  This is because, just as in electrodynamics,
its conjugate momentum is weakly zero and first class and hence eliminates
$A_0$ as an observable.

The constraint (2.3) is somewhat loosely stated.  It is important to formulate
it more accurately by first smearing it
with a suitable class of ``test'' functions
$\Lambda^{(0)}$.  Thus we write,
instead of (2.3),
\be
g(\Lambda^{(0)}) : \; =\frac{k}{2\pi}
 \int_{D} \Lambda^{(0)}(x) dA(x) \approx 0 \; .
\ee
It remains to state the space
${\cal T}^{(0)}$ of test functions $\Lambda^{(0)}$.
For this purpose, we recall that a functional on phase space can be re
lied on
to generate well defined canonical transformations only if it is
differentiable.  The meaning and implications of this remark can be illustrated
here by varying $g(\Lambda^{(0)})$ with respect to $A_{\mu}$:
\be
\delta g (\Lambda^{(0)}) =
\frac{k}{2\pi}\left ( \int_{\partial D}\Lambda^{(0)} \delta A -
\int_{D}d \Lambda^{(0)} \delta A \right ).
\ee
By definition, $g(\Lambda^{(0)})$ is differentiable in $A$ only if the boundary
term
-- the first term -- in (2.5) is zero.  We do not wish to constrain the phase
space by legislating $\delta A$ itself to be zero on $\partial D$ to achieve
this goal.  This is because we have a vital interest in regarding fluctuations
of $A$ on $\partial D$ as dynamical and hence allowing canonical
transformations
which change boundary values of $A$.  We are thus led to the following
condition
on functions $\Lambda^{(0)}$ in ${\cal T}^{(0)}$:
\be
\Lambda^{(0)} \mid_{\partial D} = 0 \;.
\ee

 It is useful to illustrate the sort of troubles we will encounter if (2.6) is
dropped.  Consider

\be
q(\Lambda) =  \frac { k } {2\pi}
\int_{D} d\Lambda A
\ee
It is perfectly differentiable in $A$ even if the function $\Lambda$ is nonzero
on $\partial D$.  It creates fluctuations
$$\delta A\mid_{\partial D} = d \Lambda \mid_{\partial D}$$  of $A$ on
 $\partial D$ by canonical
transformations.  It is a function we wish to admit in our canonical approach.
Now consider its PB with $g(\Lambda^{(0)})$:
\be
\{g(\Lambda^{0}), q(\Lambda)\} = \frac{k}{2\pi} \int d^{2}x d^{2}y
\Lambda^{(0)}(x)\epsilon^{ij}\left [\partial_{j}\Lambda (y)\right ]
\left [\frac{\partial}
{\partial
x^{i}} \delta^{2}(x-y)\right ]
\ee
where $\epsilon^{ij} = \epsilon_{ij}$. This expression is quite ill
defined if $$\Lambda^{(0)}\mid_{{\partial D}}\neq 0.$$ Thus integration on $y$
first gives zero for (2.8).  But if we integrate on $x$  first,
treating derivatives of distributions by usual rules, one finds instead,
\be
- \int_{D} d\Lambda^{0}d\Lambda = -\int_{\partial D}\Lambda^{0}d\Lambda\ .
\ee
Thus consistency requires the condition (2.6).

We recall that a similar situation occurs in QED.  There, if $E_j$ is the
electric field, which is the momentum conjugate to the potential $A_j$, and
$j_0$ is the charge density, the Gauss law can be written as
\be
\bar {g}(\bar{\Lambda}^{(0)}) = \int d^{3}x
\bar{\Lambda}^{(0)}(x)\left[\partial
_{i}
E_{i}(x)-j_{0}(x)\right]\approx 0\,.
\ee

\noindent
Since
\be
\delta \bar{g} (\bar{\Lambda}^{(0)})  = \int_{r=\infty} r^{2} d\Omega
\bar{\Lambda}^{(0)}(x) \hat{x}_{i}\delta E_{i}
-\int d^{3}x \partial_{i}\bar {\Lambda}^{(0)}(x)\delta E_i(x),
r=\mid\vec{x}\mid, \hat{x} = \frac {\vec{x}}{r}
\ee
for the variation $\delta E_i$ of $E_i$, differentiability requires
\be
\bar{\Lambda}^{(0)}(x)\mid_{r=\infty}=0.
\ee
$[d\Omega$ in (2.11) is the usual volume form of the two sphere ].
The charge, or equivalently the generator of the global U(1)
transformations, incidentally is the analogue of $q(\Lambda)$.  It is got by
partial integration on the first term.  Thus let
\be
\bar{q}(\bar{\Lambda}) = -\int d^3x\partial_{i}\bar{\Lambda}(x) E_{i}(x)
-\int d^{3}x\bar\Lambda(x)j_{0}(x) \,.
\ee
This is differentiable in $E_i$ even if $\bar {\Lambda} \mid_{r=\infty}  \neq
0$
 and
generates the gauge transformation for the gauge group element
$e^{i\bar{\Lambda}}$. It need not to vanish on quantum states if
$\bar{\Lambda}\mid_{r=\infty}\neq 0$, unlike $\bar{g}(\bar{\Lambda}^{(0)})$
which is associated with the Gauss law $\bar
{g}(\bar{\Lambda}^{(0)}) \approx 0$.  But if $\bar{\Lambda}\mid_{r=\infty}=0$,
 it becomes
the Gauss law on partial integration and annihilates all physical states.  It
follows that if $(\bar{\Lambda}_{1}-\bar{\Lambda}_{2}) \mid_{r=\infty} =0$,
then
$\bar {q} (\bar{\Lambda}_1) = \bar{q} (\bar {\Lambda}_{2})$ on physical states
which are thus sensitive only to the boundary values of test functions.  The
nature of their response determines their charge.  The conventional electric
charge of QED is $\bar {q}({\bf \bar{ 1}})$ where $\bar {\bf 1}$ is the
constant
function with value  1.

The constraints $g(\Lambda^{(0)})$ are first class since
\begin{eqnarray}
\left \{g(\Lambda_{1}^{(0)}), g(\Lambda_{2}^{(0)}) \right \}
&=& \frac{k}{2\pi} \int_{D} d\Lambda_{1}^{(0)} d \Lambda_{2}^{(0)} \nonumber \\
&=& \frac {k}{2\pi} \int_{\partial D} \Lambda_{1}^{(0)} d\Lambda_{2}^{(0)}
\nonumber \\
&=& 0 \;\;\; {\rm for}~ \Lambda_{1}^{(0)},~\Lambda_{2}^{(0)} \in\;
{\cal T}^{(0)}\;.
\end{eqnarray}
$g(\Lambda^{(0)})$ generates the gauge transformation
$A \rightarrow  A+d\Lambda^{(0)}$
of $A$.

Next consider $q(\Lambda)$ where $\Lambda\mid_{\partial D}$ is not necessarily
zero.  Since

\begin{eqnarray}
\left \{q(\Lambda),g(\Lambda^{(0)}) \right \} &=&
-\frac{k}{2\pi}\int_{D} d\Lambda d\Lambda^{(0)} \nonumber \\
&=&\frac{k}{2\pi}  \int_{\partial D} \Lambda^{(0)} d\Lambda =0 \;\;
{\rm for}~\Lambda^{(0)} \in {\cal T}^{(0)},
\end{eqnarray}
they are first class or the observables of the theory.  More precisely
observables are obtained after identifying $q(\Lambda_{1})$
with $q(\Lambda_{2})$ if
$(\Lambda_{1}-\Lambda_{2}) \in {\cal T}
^{(0)}$.  For then,
$$q(\Lambda_{1})-q(\Lambda_{2}) = - g(\Lambda_{1} -\Lambda_{2})
\approx 0.$$  The functions $q(\Lambda)$
generate gauge transformations $A\rightarrow A+d\Lambda$ which
do not necessarily vanish
on $\partial D$.

It may be remarked that the expression for $q(\Lambda)$ is obtained from
$g(\Lambda^{(0)})$ after a partial integration and a subsequent substitution of
$\Lambda$ for $\Lambda^{(0)}$.  It too  generates gauge transformations like
$g(\Lambda^{(0)})$, but the test function space for the two are different.  The
pair $q(\Lambda),g(\Lambda^{(0)})$ thus resemble the pair
$\bar{q}({\bar\Lambda}),\bar{g}(\bar{\Lambda}^{(0)})$ in QED.
The resemblance suggests
that we think of $q(\Lambda)$ as akin to the generator of a global symmetry
transformation. It is natural to do so for another reason as well:
the Hamiltonian is a constraint for a first order Lagrangian such as the one we
have here, and for this Hamiltonian, $q(\Lambda)$ is a constant of motion.

In quantum gravity, for asymptotically flat spatial slices, it is often the
practice to include a surface term in the Hamiltonian which would otherwise
have been a constraint and led to
trivial evolution \cite{adm}.  However, we know
of no natural choice of such a surface term, except zero, for the CS theory.

The PB's of $q(\Lambda)$'s are easy to compute:
\be
\{q(\Lambda_{1}),q(\Lambda_{2})\} = \frac {k}{2\pi} \int_{D} d\Lambda_{1}
d\Lambda_{2} =
\frac {k}{2\pi} \int_{\partial D} \Lambda_{1} d \Lambda_{2} \;.
\ee
Remembering that the observables are characterized by boundary values of test
functions, (2.16) shows that the observables generate a U(1) Kac-Moody algebra
\cite{godd} localized on $\partial D$. It is a Kac-Moody algebra for ``zero
momentum'' or ``charge''. For if $\Lambda \mid_{\partial D}$ is  a constant,
it can be
extended as a constant
function to all of $D$ and then $q(\Lambda) = 0$.  The central charges and
hence
the representation of (2.16) are different for $k>0$ and $k<0$, a fact which
reflects parity violation by the action $S$.

Let $\theta$ (mod $2\pi$) be the coordinate on $\partial D$ and $\phi$
a free massless
scalar field moving with speed $v$ on $\partial D$ and obeying the equal time
PB's
\be
\{\phi(\theta), \dot {\phi}(\theta^\prime)\}=\delta(\theta-\theta^\prime)\;.
\ee
If $\mu_{i}$ are test functions on $\partial D$ and
$\partial_{\pm}=\partial_{x^{0}}\pm v \partial_{\theta}$,
then
\be
\left \{\frac{1}{v} \int \mu_{1} (\theta)\partial_{\pm}
\phi(\theta),\frac{1}{v}\int
\mu_{2}(\theta)\partial_{\pm}\phi(\theta) \right \}=\pm2 \int \mu_{1}(\theta)
d \mu_{2}(\theta) ,
\ee
the remaining PB's being zero. Also $\partial_{\mp}\partial_{\pm}\phi = 0$.
Thus
the algebra of observables is isomorphic to that generated by the left moving
$\partial_{+} \phi$ or the right moving $\partial_{-}\phi$.

The CS interaction is invariant under diffeos of $D$.  An infinitesimal
generator of a diffeo with vector field $V^{(0)}$ is \cite{diff}
\be
\delta(V^{(0)})= -\frac {k}{2\pi} \int_{D}V^{(0)i}A_{i}dA.
\ee
The differentiability of $\delta(V^{(0)})$ imposes the constraint
\be
V^{(0)}\mid_{\partial D}=0\;.
\ee
Hence, in view of (2.4) as well, we have the result
\be
\delta(V^{(0)})=-\frac{k}{4\pi}\int_{D}A{\cal L}_{V^{(0)}}A \approx 0\;
\ee
where ${\cal L}_{V^{(0)}}A$ denotes the Lie derivative
of the one form $A$ with respect to  the vector field $V^{(0)}$ and is given
by $$({\cal L}_{V^{(0)}}A)_{i}
= \partial_{j}A_{i}V^{(0)j} + A_{j}\partial_{i}V^{(0)j}.$$

Next, suppose that $V$ is a vector field on $D$ which on $\partial D$ is
tangent
 to
$\partial D$,
\be
V^{i}\mid_{\partial D}(\theta) = \epsilon (\theta)
\left( \frac {\partial x^i}{\partial\theta} \right ) \mid_{\partial D} ,
\ee
$\epsilon$ being any function on $\partial D$ and $x^{i}\mid_{\partial D}$ the
restriction of $x^{i}$ to $\partial D$. $ V$ thus generates a diffeo
mapping $\partial D$ to $\partial D$. Consider next
\begin{eqnarray}
l(V)&=&\frac{k}{2\pi}\left(~\frac{1}{2}\int_{D} d(V^{i}A_{i}A) - \int_{D}
V^{i}A_{i}dA \right)~ \nonumber \\
&=&-\frac{k}{4\pi}\int_{D}A{\cal L}_{V}A .
\end{eqnarray}
Simple calculations show that $l(V)$ is differentiable in $A$ even if
$\epsilon(\theta) \neq 0$ and generates the infinitesimal diffeo of the vector
field $V$. We show in
the next Section that it is, in fact, related to $q(\Lambda)$'s by the Sugawara
construction.

The expression (2.23) for the diffeo generators of observables seems to be new.

As final points of this Section, note that
\be
\{l(V),g(\Lambda^{(0)})\}= g(V^{i}\partial_{i}\Lambda^{(0)})
=g({\cal L}_{V}\Lambda^{(0)})\approx 0\;,
\ee
\be
\{l(V),q(\Lambda)\}=q(V^{i}\partial_{i}\Lambda)=q({\cal L}_{V}\Lambda),
\ee
\be
\{l(V),l(W)\} =l({\cal L}_{V}W)
\ee
where ${\cal L}_{V}W$ denotes
the Lie derivative of the vector field $W$ with respect to the vector
field $V$ and is given
by $$({\cal L}_{V}W)^{i} =  V^{j}\partial_{j}W^{i}-W^{j}\partial_{j}V^{i}.$$
$l(V)$ are first class in view of (2.24).
Further, after the imposition of constraints, they are entirely characterized
by $\epsilon(\theta)$, the equivalence class of $l(V)$ with the same
$\epsilon(\theta)$ defining an observable.

\sxn{QUANTIZATION}

\indent
Our strategy for quantization relies on the observation that if
$$\Lambda\mid_{\partial D}(\theta)=e^{iN\theta}~,$$ then the PB's (2.16)
become those of
creation and annihilation operators.  These latter can be identified with the
similar operators of the chiral fields $\partial_{\pm}\phi$.

Thus let $\Lambda_{N}$ be any function on  $D$ with boundary value
$e^{iN\theta}$:
\be
\Lambda_{N}\mid_{\partial D}(\theta)= e^{iN\theta},\;\;N \in {\bf Z}\;.
\ee
These $\Lambda_{N}$'s exist.  All $q(\Lambda_{N})$ with the same
$\Lambda_{N}\mid_{\partial D}$ are weakly equal and define the same observable.
Let $\langle q(\Lambda_{N})\rangle$ be this equivalence class
and $q_{N}$ any member
thereof. [$q_{N}$ can also be regarded as the equivalence class itself.] Their
PB's follow from (2.16):
\be
\{q_{N},q_{M}\} = - i N k \delta_{N+M,0} \;.
\ee
The $q_{N}$'s are the CS constructions of the Fourier modes of a massless
chiral scalar field on $S^{1}$.

The CS construction of the diffeo generators $l_{N}$ on $\partial D$ (the
classical analogues of the Virasoro generators) are similar.  Thus let

 $$< l (V_{N})>$$ be the equivalence class of $l (V_{N})$
defined by the constraint
\be
V^{i}_{N} \mid_{\partial D} = e^{iN\theta}~
\left (\frac {\partial x^i}{\partial \theta}\right)\mid_{\partial D},
\sp N \in {\bf Z},
\ee
$(x^{1},~ x^{2})\mid_{\partial D}(\theta)$ being chosen to be
$R(\cos \theta,\sin
\theta)$ where $R$ is the radius of $D$.
Let $l_N$ be any member of $$<l(V_N)>~.$$ It can be
verified that
\be
\{l_N,q_M\} = i M q_{N+M} \;,
\ee
\be
\{l_N,l_M\} = - i(N-M)~ l_{N+M} \;.
\ee
These PB's are independent of the choice of the representatives from their
respective equivalence classes.  Equations (3.2), (3.4) and (3.5) define the
semidirect product of the Kac-Moody algebra and the Witt algebra (Virasoro
algebra without the central term) in its classical version.

We next show that
\be
l_N \approx \frac {1}{2k} \sum_{M}q_{M}~q_{N-M}
\ee
which is the classical version of the Sugawara construction \cite{godd}.

For convenience, let us introduce polar coordinates $r,\theta$ on $D$ ( with
$r = R$ on $\partial D$ ) and write the fields and test functions
as functions of polar coordinates. It is then clear that
\be
l_N \equiv l(V_N) = \frac{k}{4 \pi}\int_{\partial D}d\theta e^{iN\theta}
A^{2}_{\theta}(R,\theta) - \frac{k}{2\pi}
\int_{D}V^{l}_{N}(r,\theta)A_{l}(r,\theta)dA(r,\theta)
\ee
where $ A = A_{r}dr + A{\theta}d\theta.$

Let us next make the choice
\be
 e^{iM\theta}\lambda (r),~ \lambda(0)=0 \;,~~\lambda(R)=1
\ee
for $\Lambda_{M}$.
Then
\be
q_M = q(e^{iM\theta}\lambda(r)).\ee
Integrating (3.9) by parts, we get
\be
q_M = \frac {k}{2 \pi}\left (
\int_{\partial D} d\theta e^{iM\theta}A_{\theta}(R,\theta)
- \int_{D} dr d{\theta} \lambda (r) e^{iM\theta} F_{r \theta}(r,\theta) \right
)
\ee
where $F_{r \theta}$ is defined by $dA = F_{r\theta}dr \wedge
d\theta$.
Therefore
\begin{eqnarray}
\frac{1}{2k} \sum_{M} q_{M}q_{N-M}~=
&+&\frac{k}{4\pi}\int_{\partial D} d\theta  e^{iN\theta} A^2_{\theta}(R,\theta)
 \nonumber \\
&-&\frac{k}{2 \pi}\int_{D} dr d\theta  e^{iN\theta}\lambda (r)
 A_{\theta}(R, \theta) F_{r \theta}(r, \theta)
\nonumber \\
&+&\frac{k}{4 \pi}\int_{D} dr d\theta dr^\prime \lambda (r) \lambda
(r^{\prime})e^{iN\theta}
F_{r \theta}(r, \theta) F_{r \theta}(r^{\prime}, \theta)
\end{eqnarray}
where the completeness relation
$$\sum_N e^{iN(\theta-\theta^\prime)} = 2\pi \delta (\theta-\theta^\prime)$$
has been used.

The test functions for the Gauss law in the last term in (3.11)
involves $F_{r \theta}$ itself. We therefore interpret it to be zero and get
\be
\frac{1}{2k} \sum_{M} q_{M}q_{N-M} \approx
 \frac{k}{4\pi} \int_{\partial D} e^{iN\theta} A^2_{\theta}(R,\theta) d
\theta - \frac{k}{2\pi}\int_{D}dr d\theta
e^{iN\theta}\lambda(r)A_{\theta}(R,\theta)F_{r \theta}(r,\theta).
\ee

 Now in view of  (3.3) and (3.8), it is clear that
\be
V^{l}_{N}(r,\theta)A_{l}(r,\theta) - e^{iN\theta}\lambda(r)A_{\theta}(R,\theta)
= 0 ~~~ {\rm on}~~ \partial D.
\ee
Therefore $$l_N \approx \frac {1}{2k} \sum_{M} q_{M}q_{N-M} $$
which proves (3.6).

We can now proceed to quantum field theory. Let ${\cal
G}(\Lambda^{(0)}),Q(\Lambda_N),Q_N$ and $L_N$ denote the quantum operators for
$g(\Lambda^{(0)}), q(\Lambda_N),q_N$ and $l_N$.  We then impose the constraint
\be
{\cal G}(\Lambda^{(0)}) | \cdot \rangle = 0
\ee
on all quantum states.  It is an expression of their gauge invariance.  Because
of this equation, $Q(\Lambda_N)$ and $Q(\Lambda ^\prime_N)$ have the same
action
on the states if $\Lambda_N$ and $\Lambda ^\prime_N$ have the same boundary
values.  We can hence write
\be
Q_N | \cdot \rangle = Q(\Lambda_N) | \cdot \rangle \;.
\ee
Here, in view of (3.2), the commutator brackets of $Q_N$ are
\be
[Q_N,Q_M] = N k \delta_{N+M,0}\;.
\ee

Thus if $k>0 \;\;(k<0),Q_N$ for $N>0 \;(N<0)$ are annihilation operators
( upto a normalization ) and
$Q_{-N}$ creation operators.  The ``vacuum'' $| 0 >$ can therefore be defined
by
\be
Q_N \mid 0> = 0 \;\;{\rm if}~ Nk>0\;.
\ee
The excitations are got by applying $Q_{-N}$ to the vacuum.

The quantum Virasoro generators are the normal ordered forms of their classical
expression \cite{godd} :
\be
L_N =  \frac {1}{2k}:~\sum_{M}Q_M Q_{N-M}:
\ee
They generate the Virasoro algebra for central charge $c=1$ ~:
\be
\left [L_N,L_M \right ] = (N-M)L_{N+M}
+ \frac {c}{12} (N^3-N) \delta_{N+M,0}~,~c=1 \;\;.
\ee

When the spatial slice is a disc, the observables are all given by $Q_N$ and
our quantization is complete.  When it is not simply connected, however, there
are further observables associated with the holonomies of the connection $A$
and
they affect quantization.  We will not examine quantization for nonsimply
connected spatial slices here.

The CS interaction does not fix the speed $v$ of the scalar field in (2.18) and
so its Hamiltonian, a point previously emphasized by
Frohlich and
Kerler \cite{zhang} and Frohlich and Zee \cite{blok}.
This is but reasonable.  For if we could fix $v$, the Hamiltonian H for $\phi$
could naturally be taken to be the one for a free massless chiral scalar field
moving with speed $v$.  It could then be used to evolve the CS observables
using
the correspondence of this field and the former.  But we have seen that no
natural nonzero Hamiltonian exists for the CS system.  It is thus satisfying
that we can not fix $v$ and hence a nonzero H.

In the context of Fractional Quantum Hall Effect, the following generalization
of the CS action has become of interest \cite{blok}:
\be
S ^\prime = \frac{k}{4\pi} {\cal K}_{IJ} \int_{D \times {\bf R}^{1}} A^{(I)}
dA^
{(J)}.
\ee
Here the sum on $I,J$ is from $1$ to $F,~A^{(I)}$ is associated with the
current
$j^{(I)}$ in the $I^{th}$ Landau level and
${\cal K}$ is a certain invertible symmetric
real $F \times F$ matrix . By way of further illustration
of our approach to
quantization,  we now outline the quantization of (3.20) on $D \times {\bf
R}^{1}$.

The phase space of (3.20) is described by the PB's
\be
\left \{ A^{(I)}_{i} (x), A^{(J)}_j(y) \right \} = \epsilon_{ij}\frac{2\pi}{k}
{\cal K}^{-1}_{IJ}\delta^{2}(x-y),~~x^0=y^0
\ee
and the first class constraints
\be
g^{(I)}(\Lambda^{(0)}) =
\frac{k}{2\pi} \int_D \Lambda^{(0)}dA^{(I)}\approx 0\ , ~~ \Lambda^{(0)} \in
{\cal T}^{(0)}
\;.
\ee
with zero PB's.

The observables are obtained from the first class variables
\be
q^{(I)}(\Lambda) = \frac{k}{2\pi} \int_D d\Lambda A^{(I)}
\ee
after identifying $q^{(I)}(\Lambda)$ with $q^{(I)}(\Lambda^\prime)$ if
$(\Lambda-\Lambda^\prime)\mid_{\partial D}=0$. The PB's of $q^{(I)}$'s are
\be
\left \{q^{(I)}(\Lambda^{(I)}_{1}), q^{(J)}(\Lambda^{(J)}_{2})\right \} =
\frac{k}{2\pi}{\cal K}^{-1}_{IJ} \int_{\partial D}
\Lambda^{(I)}_{1}d\Lambda^{(J)}_{2}.
\ee

Choose a $\Lambda^{(I)}_{N}$ by the requirement $\Lambda^{(I)}_N \mid_{\partial
D}(\theta)= e^{iN\theta}$ and let $q^{(I)}_N$ be any member of the equivalence
class $<q^{(I)}(\Lambda^{(I)}_N)>$ characterized by such $\Lambda^{(I)}_N$.
Then
\be
\left \{q^{(I)}_N,q^{(J)}_M \right \}= -i {\cal K}^{-1}_{IJ} Nk\delta_{N+M,0}~.
\ee

As ${\cal K}^{-1}_{IJ}$ is real symmetric, it can be diagonalized by a real
orthogonal
transformation $M$ and has real eigenvalues $\lambda_{\rho} ~(\rho=1,2,...,F)$.
As ${\cal K}^{-1}_{IJ}$ is invertible, $\lambda_{\rho} \neq 0$. Setting
\be
q_N(\rho)=M_{\rho I}q^{(I)}_N
\ee
we have
\be
\left \{q_N(\rho), q_M(\sigma)\right \} = -i \lambda_\rho Nk
\delta_{\rho\sigma}
\delta_{N+M,0}~~.
\ee
(3.27) is readily quantized. If $Q_N(\rho)$ is the quantum operator for
$q_N(\rho)$,
\be
\left [Q_N(\rho),Q_M(\sigma)\right] =  \lambda_{\rho} Nk
\delta_{\rho\sigma}\delta_{N+M,0}~.
\ee
(3.28) describes F harmonic oscillators or edge currents.  Their chirality, or
the chirality of the corresponding massless scalar fields, is governed by the
sign of $\lambda_\rho$.

The classical diffeo generators for the independent oscillators $q_N(\rho)$ and
their quantum versions can be written down using the foregoing discussion.  The
latter form F commuting Virasoro algebras, all for central charge 1.

\sxn{THE NONABELIAN CHERN-SIMONS ACTION}

\indent
Let G be a compact simple group with Lie algebra ${\underline G}$.  Let
$\gamma$
 be a faithful
representation of ${\underline G}$. Choose a hermitian basis
$\{T_{\alpha}\}$ for
$\gamma$ (more precisely $i\gamma$) with normalization
$Tr~T_{\alpha}T_{\beta}=\delta_{\alpha\beta}$.
Let $A_{\mu}$
define an antihermitean connection for G with values in $\gamma$.  We define
the real field $A^{\alpha}_{\mu}$ by
$A_{\mu}=iA^{\alpha}_{\mu}T_{\alpha}$.  With these conventions, the
Chern-Simons action for $A_{\mu}$ on $D\times {\bf R}^{1}$ is
\be
S=-\frac {k}{4\pi} \int_{D\times {\bf R}^{1}} Tr \left[AdA+\frac{2}{3}
A^3\right],~~A=A_{\mu}dx^{\mu}
\ee
where the constant $k$ can assume only quantized values for well known reasons.
If $G=SU(N)$ and $\gamma$ the Lie algebra of its defining representation, then
$k\in {\bf Z}$.

Much as for the Abelian problem, the phase space for (4.1) is described by the
PB's
\be
\left \{A^{\alpha}_{i}(x),A^{\beta}_{j}(y)\right\} =
\delta_{\alpha\beta}~\epsilon_{ij}~\frac {2\pi}{k}~\delta^2(x-y),~~~x^0=y^0
\ee
and the Gauss law
\be
g({\Lambda}^{(0)}) = -\frac {k}{2\pi} \int_{D} Tr~\left \{ {
\Lambda}^{(0)}(dA+A^{2} )\right \} =-\frac{k}{2\pi} \int_{D} Tr~({
\Lambda^{(0)}F)}\approx 0
\ee
where $F=F_{ij}dx^{i} dx^{j}$ is the curvature of $A$,
${\Lambda}^{(0)}=i\Lambda^{(0)\alpha}T_{\alpha}$ and
$\Lambda^{(0)\alpha}\in {\cal T}^{(0)}$.  This test function
space for ${\Lambda}^{(0)}$ ensures that
$g({\Lambda}^{(0)})$ is differentiable in $A^{\alpha}_i$.  The PB's
between $g$'s are
\be
\left\{g({\Lambda_{1}}^{(0)}), g{(\Lambda_{2}}^{(0)})\right\} =
g([{\Lambda_{1}}^{(0)},{\Lambda_{2}}^{(0)}])-\frac
{k}{2\pi} \int_{\partial D} Tr~{\Lambda_{1}^{0}d\Lambda_{2}}^{(0)} =
g([{\Lambda_{1}}^{(0)},{\Lambda_{2}}^{(0)}])
\ee
so that they are first class constraints.

Next define
\be
q({\Lambda})= \frac {k}{2\pi} \int_{D}
Tr~(-d{\Lambda}A+{\Lambda}A^{2}),~~~{\Lambda}=i\Lambda^
{\alpha}T_{\alpha}~.
\ee
It is differentiable in $A^{\alpha}_{i}$ even if
${\Lambda}|_{\partial D}\neq 0$.  But if
${\Lambda}|_{\partial D}$ is zero, it is equal to the Gauss law
$g({\Lambda})$.  Further, $q({\Lambda})$ is first class for
any choice of ${\Lambda}$ since

\be
\left \{ q(\Lambda), g({\Lambda}^{(0)})\right \} =
-g([{\Lambda},{\Lambda}^{(0)}])\approx 0~.
\ee
Thus (with ${\Lambda}|_{\partial D}$ free),
$q({\Lambda})$'s define observables, the latter being the same if
their test functions are equal on $\partial D$.

\newcommand{\uL}{\Lambda}
\newcommand{\uLs}{\Lambda^{(0)}}
The PB's of $q(\Lambda)$'s are
\be
\left\{q(\Lambda_{1}),q({\Lambda_{2}})\right\}=-q([\Lambda_{1},
{\Lambda_{2}}])-\frac{k}{2\pi} \int_{\partial D}Tr~\left(\Lambda_{1} d
{\Lambda_{2}}\right)
\ee
which can be recognized as a Kac-Moody algebra for observables.

The diffeo generators can also be constructed following Section 3. The
generators of diffeos which keep $\partial D$ fixed and vanish weakly are
\be
\delta(V^{(0)}) = \frac {k}{2\pi} \int_{D} V^{(0)i}~TrA_iF,~~~ V^{(0)i}|_
{\partial D}=0~,
\ee
while those generators which also perform diffeos of $\partial D$ are
\begin{eqnarray}
l(V) &=& \frac {k}{2\pi}~\left(~ \int_{D} V^{i}~TrA_iF-\frac{1}{2}\int_{D}d(V^i
{}~TrA_iA)\right) \nonumber \\
&=& \frac{k}{4 \pi}\int_{D}TrA{\cal L}_{V}A
\end{eqnarray}
where $V^{i}|_{\partial
D}(\theta)=\epsilon(\theta)\left({\frac{\partial x^{i}}{\partial
\theta}}\right)
{\mid_{\partial D}}$.
The PB's involving $l(V)$ are patterned after (2.24--2.26):
\be
\left\{l(V ),g(\uLs)\right\} = g ( V^i\partial_i \uLs ) = g ({\cal
L}_{V}\Lambda^{(0)}) \approx 0~,
\ee
\be
\left\{l(V),q(\Lambda)\right\} = q(V^i\partial_i \uL)~=q ({\cal
L}_{V}\Lambda)~,
\ee
\be
\left\{ l(V),l(W)\right\}= l ({\cal L}_VW)~.
\ee
We can now conclude that $l(V)$ are first class and define observables, all $V$
with the same $\epsilon (\theta)$ leading to the same observable.

\newcommand{\Lan}{(\Lambda^{\alpha}_{N})}
Let $\Lambda^{\alpha}_N$ be any test function with the feature
$\Lambda^{\alpha}_{N}|_{\partial D}=e^{iN\theta} T_{\alpha}$ and let
$V^i_N$ be defined following Section 3.  As in that Section, let us call the
set of first class variables weakly equal to
$q (i\Lambda^{\alpha}_{N}T_{\alpha})$ and $l(V_N)$ by $\langle
q (i\Lambda^{\alpha}_{N}T_{\alpha})\rangle$
and $\langle l(V_N)\rangle$. [ Here there is no sum over $\alpha$ in
$i\Lambda^{\alpha}_{N}T_{\alpha}$].
Let
$q^{\alpha}_N$ and $l_N$ be any member each from these sets.  Their PB's are
\be
\left\{ q^{\alpha}_N, q^{\beta}_M\right\}
\approx f_{\alpha\beta\gamma} q^{\gamma}_{N+M}-
iNk  \delta_{N+M,0}~\delta_{\alpha\beta}~,
\ee
\be
\left\{l_N,q^\alpha_M\right\} \approx iM q^{\alpha}_{N+M}~,
\ee
\be
\left\{l_N,l_M\right\} \approx -i(N-M) l_{N+M},
\ee
$f_{\alpha\beta\gamma}$ being defined by $[T_\alpha,T_\beta]=
if_{\alpha\beta\gamma}T_\gamma$. Furthermore, as in Section 3,
\be
l_N \approx \frac {1}{2k} \sum_{M,\alpha}q^{\alpha}_{M}q^{\alpha}_{N-M}~.
\ee

We next go to quantum field theory.  In quantum theory, the operators for
$g(\uLs),q^\alpha_N$ and $l_N$ are denoted by ${\cal G}(\uLs),
Q^\alpha_N, L_N$ and all states are subjected to the Gauss law
\be
{
\cal G}(\uLs)|\cdot>= 0~.
\ee
As a consequence, all the weak equalitites can be regarded as strong for the
quantum operators.  We are thus dealing with a Kac-Moody algebra for a certain
level \cite{godd}. A suitable highest weight
representation for it can be constructed
in the usual way \cite{godd}, thereby defining the quantum theory.
The expression for
the Virasoro generators normalized to fulfill the commutation relations
(3.16) is not the normal ordered version of (4.16), but as is well known, it is
\be
L_N=\frac {1}{2k+c_{V}} \sum_{M,\alpha}~:~Q^\alpha_M Q^\alpha_{N-M}~:~,
\ee
($c_{V}$ being the quadratic Casimir operator in the adjoint representation).
 The central charge $c$ now is not
of course 1, but rather,
\be
c= \frac {2k\dim {\underline G}} {2k+c_{V}}~~ ,{\rm dim}{\underline G}
\equiv {\rm dimension~ of~}{\underline G }.
\ee
These results about the Kac-Moody and Virasoro algebras are explained in
ref. 2.\\

{\bf Acknowledgement}

We have been supported during the course of this work as follows:~1) A.~P.~B.,
G.~B.,~K.~S.~G.~ by the Department of Energy, USA, under contract number
DE-FG-0
2-85ER
-40231, and A.~S.~by the Department of Energy, USA under contract
number DE-FG05-84ER40141;
2) A.~P.~B.~and A.~S.~ by INFN, Italy [at Dipartimento di
Scienze Fisiche, Universit{\`a} Di Napoli]; 3) G.~B.~ by the Dipartimento
di Scienze Fisiche, Universit{\`a} di Napoli.  The authors wish to thank the
 group in Naples and Giuseppe Marmo, in particular, for their hospitality while
this
work was in progress. They also wish to thank Paulo Teotonio for very helpful
comments and especially for showing us how to write equations (2.21),
(2.23) and (4.9)
in their final nice forms involving Lie derivatives.


\newpage

\begin{thebibliography}{99}

\bibitem{witt}
E. Witten,  Commun.~Math.~Phys.~{\bf 121},~351~(1989).

\bibitem{godd} For a review, see P. Goddard and D. Olive,
Int.~J.~Mod.~Phys.~{\bf A1}, 303 (1986).

\bibitem{bos} M. Bos and V. P. Nair, Int. J. Mod. Phys. {\bf A5}, 959 (1990);
Phys. Lett. {\bf B223}, 61 (1989); ; T. R. Ramadas, Comm. Math. Phys. {\bf
128}, 421 (1990); D. Boyanovsky, E. T. Newman and C. Rovelli, University of
Pittsburgh preprint PITT-91-14 (1991).

\bibitem{bal} A. P. Balachandran, M. Bourdeau
and S. Jo,  Mod. Phys. Lett. {\bf A4}, 1923 (1989); Int. J. Mod. Phys.
{\bf A5}, 2423 (1990) [Erratum : ibid. 3461, (1990); K. Gupta and A. Stern,
Phys. Rev. {\bf D44}, 2432 (1991).

\bibitem{guad} E. Guadagnini, M. Martellini and M. Mintchev,
Nucl. Phys. {\bf B330},
575 (1990); Phys. Lett. {\bf B235}, 275 (1990);
Nucl. Phys. {\bf B336}, 581 (1990).

\bibitem{lee} T. R. Ramadas, I. M. Singer and J. Weitsman,
Comm. Math. Phys. {\bf 126}, 406 (1989);
L.~Smolin, Mod. Phys. Lett. {\bf A4}, 1091 (1989).

\bibitem{moore} G. Moore and N. Seiberg,  Phys. Lett. {\bf B220}, 422 (1989).

\bibitem{zhang} S. C. Zhang, H. Hansson, S. Kivelson , Phys.
Rev. Lett. {\bf 62}, 82 (1989);
F. Wilczek
``{\it Fractional Statistics and Anyon Superconductivity }''
(World Scientific, Singapore, 1990) and articles therein;
G. Moore and N. Read, Yale preprint YC TP-P6-90 (1990);
J. Frohlich
and T. Kerler, Nucl. Phys. {\bf B354}, 369 (1991);
G. Cristofano, G. Maiella, R. Musto and F. Nicodemi, Mod. Phys. Lett. {\bf A6},
{}~2985 (1991); M. Stone and H. W. Wyld, University of Illinois, Urbana
preprint
ILL-TH-91-21 (1991)
and references in these papers.

\bibitem{blok} B. Blok and X. G. Wen, Institute for Advanced Study,
Princeton  preprint
IASSNS-HEP-90/23 (1990); X.G.Wen, ibid, IASSNS-HEP-91/20 (1991); J. Frohlich
and A. Zee,
Institute for Theoretical Physics, Santa Barbara preprint
ITP-91-31 (1991) and references in these papers.

\bibitem{mor} See the review article of F. Wilczek in ``{\it Anomalies, Phases,
Defects,}'' edited by G. Morandi (Bibliopolis, Napoli, 1990);
A. P. Balachandran, E. Ercolessi, G. Morandi
and A. M. Srivastava ``{\it The Hubbard Model and Anyon Superconductivity}'',
Int. J. Mod. Phys. {\bf B4}, 2057 (1990) and
Lecture Notes in Physics, Vol. 138, ( World Scientific, 1990);
R. Shankar and M. Sivakumar, Mod. Phys. Lett.{\bf A6}, 2379 (1991);
P. Sharan, M. Sami and S. Alam, Jamia Millia Islamia, New Delhi preprint
(1991);
issue on
`` {\it Fractional Statistics in Action}'', edited by F. Wilczek, Int. J. Mod.
Phys {\bf B} (1991) (in press).


\bibitem{marmo} Cf. A. P. Balachandran, G. Marmo, B.- S. Skagerstam and A.
Stern
``{\it Classical Topology and Quantum States}'' (World Scientific, Singapore,
1991),Part I.

\bibitem{buch} D. Buchholz and K. Fredenhagen,  Commun. Math. Phys. {\bf
84}, 1 (1982) ; K. Fredenhagen, K. H. Rehren and B. Schroer,  Commun. Math.
Phys. {\bf 125}, 201 (1989) and preprints (1990); K. Fredenhagen,
Proceedings of
the Lake Tahoe Meeting (1990); J. Frohlich and P. A. Marchetti,
Universit{\`a} di Padova preprint
DFPD/TH/10/90 (1990) and references therein.

\bibitem{bim} A. P. Balachandran, G. Bimonte, K. S. Gupta and
A. Stern,``Conformal Edge Currents in Chern-Simons Theories,'' talk presented
by K. S. Gupta at the Montreal-Rochester-Syracuse-Toronto meeting, Rochester
(1991), Syracuse University preprint SU-4228-
477, INFN-NA-IV-91/22, UAHEP 915(1991) and Proceedings of the Thirteenth Annual
Montreal-Rochester-Syracuse-Toronto meeting (1991) page 25.

\newpage

\bibitem{adm} See the article of R. Arnowitt, S. Deser and C. Misner in
``{\it Gravitation : An Introduction to Current Research}'' (Wiley, 1962).

\bibitem{diff} E. Witten, Nucl. Phys. {\bf B311}, 46 (1988).


\end{thebibliography}
\end{document}